\documentclass{article}
\usepackage{graphicx}

\begin{document}

\title{Novel Aharonov-Bohm-like effect: Detectability of the vector
potential in a solenoidal configuration with a ferromagnetic core covered by
superconducting lead, and surrounded by a thin cylindrical shell of aluminum}
\date{White paper of June 22, 2012}
\author{R.Y. Chiao (rchiao@ucmerced.edu)}
\maketitle

\abstract{The flux as measured by the Josephson effect in a SQUID-like configuration with a ferromagnetic core inserted into its center, is shown to be sensitive to the vector potential arising from the central ferromagnetic core, even when the core is covered with a superconducting material that prevents any magnetic field lines from ever reaching the perimeter of the SQUID-like configuration. This leads to a macroscopic, Aharonov-Bohm-like effect that is observable in an asymmetric hysteresis loop in the response of the SQUID-like configuration to an externally applied magnetic field.}\\

\begin{figure}
\includegraphics[width=4.75in]{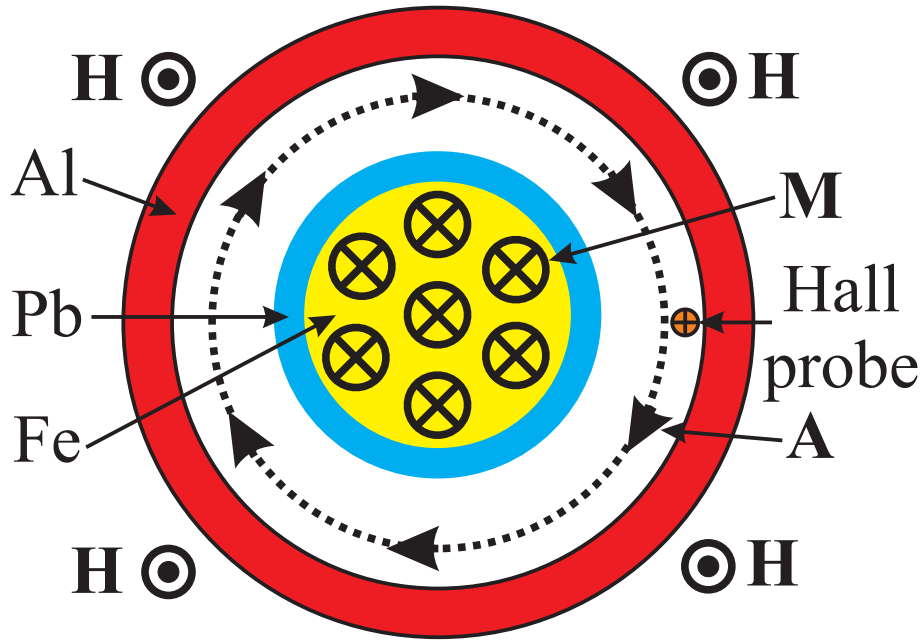}
\caption{A solenoidal configuration
consists of an infinitely long ferromagnetic core (yellow) coated by a thick
layer of lead (blue), at the center of a thin cylindrical shell of aluminum
(red). $\bf{H}$ is an externally applied magnetic field, $\bf{M}$
the magnetization of the ferromagnetic core, and $\bf{A}$ the vector
potential. As the temperature is steadily lowered, the lead (blue) coating
first becomes superconducting, and totally confines the ferromagnetic flux
to the interior of the core via the Meissner effect, so that the aluminum
ring detects no $\bf{B}$ field arising from the core, but detects the $%
\bf{A}$ field. As the temperature is further lowered, the aluminum (red)
ring becomes superconducting, and acquires a persistent current which
measures the core's $\bf{A}$ field\ via a SQUID-like flux trapping
effect.}
\end{figure}

In our ongoing experiments \cite{ChiaoMayArXiv}, we are using a ferromagnetic torus coated with
a superconductor (lead) \cite{Tonomura}, which is topologically linked with another
superconducing (SC) ring made out of a different material (aluminum), which
has a lower critical temperature. However, to understand this experiment more easily, let us first consider the simpler cylindrically-symmetric configuration of Figure 1, which depicts a straight, infinitely long solenoidal configuration consisting of a
cylindrical ferromagnetic core (yellow) possessing a permanent magnetization $\bf{M%
}$, which is covered with a thick coating of SC lead (blue), in order to
prevent, due to the Meissner effect, any stray magnetic flux lines from
escaping from the core. This core in turn is placed at the center of a thin
cylindrical shell of SC aluminum (red), which has a lower critical
temperature than that of SC lead (blue).

Let us further assume that the temperature has been cooled to near absolute
zero, and that we are slowly applying a uniform magnetic field $\bf{H}$\
onto the entire assembly starting from zero field. The magnetostatic flux
filling the ferromagnetic core due to its permanent magnetization will give
rise to a static contribution to the circular lines of the vector potential $%
\bf{A}$, as illustrated by the dashed circle in Figure 1. However, note
that due to the Meissner effect arising from the thick lead coating, the
magnetostatic field $\bf{B}$ arising from the ferromagnetic core
vanishes in the spatial region in which the cylindrical shell of aluminum
material resides, so that the electrons in this shell never experience any
classical forces. This is true both above and below the critical temperature
of the aluminum shell. Nevertheless, like in the Aharonov-Bohm effect, there
exists a nonlocal quantum influence\ of the electrons within the
ferromagnetic core upon the electrons within the superconducting shell,
since persistent currents within the core can induce persistent currents
within the shell that are proportional to the size of the vector potential
arising from the ferromagnetic core.

The probability current density in quantum mechanics for a single neutral
particle of mass $m$ moving through the vacuum is given by%
\begin{eqnarray}
{\bf j} &=&\frac{\hbar }{2mi}\left( \psi ^{\ast }{\bf \nabla }\psi
-\psi {\bf \nabla }\psi ^{\ast }\right)  \nonumber \\
&=&\frac{1}{2m}\left( \psi ^{\ast }\left( \frac{\hbar }{i}{\bf \nabla }%
\right) \psi -\psi \left( \frac{\hbar }{i}{\bf \nabla }\right) \psi
^{\ast }\right)  \nonumber \\
&=&\frac{1}{2m}\left( \psi ^{\ast }{\bf p}_{{\rm op}}\psi \right) +{\rm %
c.c.}  \label{prob curr dens}
\end{eqnarray}%
where $\psi =\psi \left( {\bf r},t\right) $ is the wavefunction evaluated
at the spacetime point $\left( {\bf r},t\right) $, and where the momentum
operator in the configuration space representation is%
\begin{equation}
{\bf p}_{{\rm op}}=\frac{\hbar }{i}{\bf \nabla }
\end{equation}%
From the time-dependent Schrodinger equation, the continuity equation%
\begin{equation}
{\bf \nabla \cdot j}+\frac{\partial \rho }{\partial t}
\label{continuity equation}
\end{equation}%
follows, in which%
\begin{equation}
\rho =\psi ^{\ast }\psi =\psi ^{\ast }\left( {\bf r},t\right) \psi \left( 
{\bf r},t\right)
\end{equation}%
is the probability density for finding the particle located at $\left( 
{\bf r},t\right) $ upon measurement. The continuity equation (\ref%
{continuity equation}) means that probability is conserved.

Let us generalize the above expression\ (\ref{prob curr dens}) for the
probability current density ${\bf j}\left( {\bf r},t\right) $\ to that
of a particle with charge $q$ moving through the vacuum in the presence of a
vector potential ${\bf A}\left( {\bf r},t\right) $, by applying the
minimal coupling rule \cite{ChiaoMayArXiv}%
\begin{equation}
{\bf p}_{{\rm op}}\rightarrow {\bf p}_{{\rm op}}-q{\bf A}
\end{equation}%
to (\ref{prob curr dens}). Then one finds that for the particle with charge $%
q$%
\begin{equation}
{\bf j}=\frac{1}{2m}\left( \psi ^{\ast }\left( \frac{\hbar }{i}{\bf %
\nabla }-q{\bf A}\right) \psi \right) +{\rm c.c.}
\label{curr dens with A}
\end{equation}

This expression for the probability current density ${\bf j}\left( 
{\bf r},t\right) $ for the charge $q$ moving through the vacuum
generalizes to an expression for the supercurrent density ${\bf J}%
_{s}\left( {\bf r},t\right) $ for Cooper pairs moving through a SC, which
can be obtained from the complex order parameter $\Psi \left( {\bf r}%
,t\right) $\ of the Ginzburg-Landau theory as follows:%
\begin{equation}
{\bf J}_{s}=\frac{1}{2m}\left( \Psi ^{\ast }\left( \frac{\hbar }{i}%
{\bf \nabla }-q{\bf A}\right) \Psi \right) +{\rm c.c.}
\label{G-L Current Density}
\end{equation}%
where $q=2e$ is the charge and $m=2m_{e}$ is the mass of the Cooper pair.

Next, let us express the complex order parameter $\Psi $ in polar form as
follows:%
\begin{equation}
\Psi =\sqrt{\rho }e^{i\phi }  \label{polar decomposition}
\end{equation}%
where%
\begin{equation}
\rho =\Psi ^{\ast }\Psi
\end{equation}%
is the probability density of Cooper pairs, and where $\phi $ is the local
phase of the Cooper pair's \textquotedblleft macroscopic
wavefunction\textquotedblright\ $\Psi $.

Since the local \emph{positive} charge density of the background ionic
lattice must be exactly locally compensated by the local \emph{negative}
charge density of the Cooper pairs in order to preserve the overall charge
neutrality of the SC material (which must be true under all conceivable
experimental circumstances), we require that%
\begin{equation}
\left\vert \rho \right\vert =\left\vert \rho _{{\rm ionic lattice}%
}\right\vert ={\rm constant}
\end{equation}%
independent of space and time. Therefore, to an extremely good
approximation, the density of Cooper pairs%
\begin{equation}
\rho \left( {\bf r},t\right) =\rho _{0}
\end{equation}%
is independent of $\left( {\bf r},t\right) $; only their phase $\phi
=\phi \left( {\bf r},t\right) $ is allowed to vary with space and time
inside the SC. Hence the polar decomposition (\ref{polar decomposition})
becomes 
\begin{equation}
\Psi \left( {\bf r},t\right) =\sqrt{\rho _{0}}e^{i\phi \left( {\bf r}%
,t\right) }
\end{equation}%
Therefore the supercurrent density (\ref{G-L Current Density}) becomes%
\begin{eqnarray}
{\bf J}_{s} &=&\frac{1}{2m}\left( \sqrt{\rho _{0}}e^{-i\phi }\left( \frac{%
\hbar }{i}\left( i{\bf \nabla }\phi \right) -q{\bf A}\right) \sqrt{%
\rho _{0}}e^{i\phi }\right) +{\rm c.c.}  \nonumber \\
&=&\frac{\rho _{0}}{2m}\left( \frac{\hbar }{i}\left( i{\bf \nabla }\phi
\right) -q{\bf A}\right) +{\rm c.c.}  \nonumber \\
&=&\frac{\rho _{0}}{m}\left( \hbar {\bf \nabla }\phi -q{\bf A}\right)
=\rho _{0}{\bf v}_{s}
\end{eqnarray}%
where ${\bf v}_{s}$ is the superfluid velocity of the Cooper pairs. Hence
the supercurrent density becomes%
\begin{equation}
{\bf J}_{s}=\rho _{0}{\bf v}_{s}=\frac{\rho _{0}}{m}\left( \hbar 
{\bf \nabla }\phi -q{\bf A}\right) 
\end{equation}%
and therefore%
\begin{equation}
m{\bf v}_{s}=\hbar {\bf \nabla }\phi -q{\bf A}
\label{three momenta}
\end{equation}%
where $m{\bf v}_{s}$ is the \emph{kinetic} momentum, $\hbar {\bf %
\nabla }\phi $ is the \emph{canonical} momentum, and $q{\bf A}$ is the 
\emph{electromagnetic} momentum of the Cooper pairs.

The concept of \emph{fluxoid} quantization follows from (\ref{three momenta}%
), by integration over a circular path inside the material of the red
cylindrical shell of Figure 1 as follows:%
\begin{equation}
m\oint {\bf v}_{s}\cdot d{\bf l}=\hbar \oint {\bf \nabla }\phi
\cdot d{\bf l}-q\oint {\bf A}\cdot d{\bf l}
\end{equation}%
Due to the single-valuedness of complex order parameter $\Psi $, it follows
that%
\begin{equation}
\oint {\bf \nabla }\phi \cdot d{\bf l}=2\pi n,{\rm  }(n=0,\pm 1,\pm
2,...)
\end{equation}%
and therefore that%
\begin{equation}
q\oint {\bf A}\cdot d{\bf l+}m\oint {\bf v}_{s}\cdot d{\bf l}%
=2\pi n\hbar =nh,{\rm  }(n=0,\pm 1,\pm 2,...)
\label{quantization of ang mom}
\end{equation}%
Let us define the \emph{fluxoid} as follows:%
\begin{equation}
\Phi _{{\rm fluxoid}}\equiv \oint \left( {\bf A+}\frac{m}{q}{\bf v}%
_{s}\right) \cdot d{\bf l}=\Phi +\frac{m}{q}\kappa 
\end{equation}%
where%
\begin{equation}
\Phi =\oint {\bf A}\cdot d{\bf l}
\end{equation}%
is the flux, and where%
\begin{equation}
\kappa =\oint {\bf v}_{s}\cdot d{\bf l}
\end{equation}%
is the circulation of the superfluid, which can become important when the
cylindrical shell becomes thin compared to the penetration depth of the SC
(here, aluminum), as happens when the applied magnetic $H$ field approaches
the critical field $H_{{\rm c}}$. It follows from (\ref{quantization of ang
mom}) that the fluxoid obeys the following quantization condition:%
\begin{equation}
\Phi _{{\rm fluxoid}}=n\frac{h}{2e}=n\Phi _{0},{\rm  }(n=0,\pm 1,\pm 2,...)
\end{equation}%
where%
\begin{equation}
\Phi _{0}=\frac{h}{2e}=2.07\times 10^{-15}{\rm  webers}
\end{equation}%
is the quantum of flux. However, if the cylindrical shell is thick compared
to the penetration depth, then the circulation $\kappa $ for the superfluid
deep inside the red ring of Figure 1 becomes negligible, and fluxoid
quantization reduces to the usual flux quantization condition, where%
\begin{equation}
\Phi _{{\rm flux}}\equiv \oint {\bf A}\cdot d{\bf l}=\Phi _{n}=n\frac{%
h}{2e}=n\Phi _{0},{\rm  }(n=0,\pm 1,\pm 2,...)
\end{equation}

\begin{figure}
\includegraphics[width=5in]{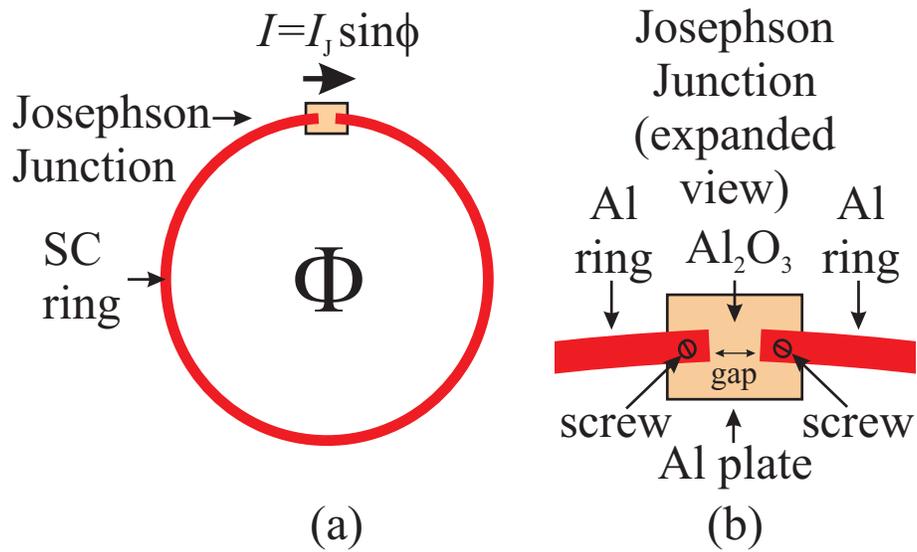}
\caption{(a) SQUID-like model for the
production of a persistent supercurrent $I$ within a SC ring (in red) by a
total enclosed flux $\Phi $, when the ring is interrupted by a Josephson
junction possessing a critical current $I_{{\rm J}}$. The self-inductance $L
$ of the ring will lead to the onset of a remnant field and hysteresis. (b)
Expanded view of the Josephson junction. An aluminum plate (pink) with an
oxidized sapphire surface (Al$_{2}$O$_{3}$), when screwed tightly onto the
gapped aluminum ring (i.e., in the form of a flat gasket with a gap in it),
makes the ring into a closed SC circuit containing a single, effective
Josephson junction in it.}
\end{figure}

Figure 2(a) depicts a SQUID-like model for the nonlinearity that occurs
within the red cylindrical shell in Figure 1 as the externally applied magnetic field
passes through the effective critical field of the SC material (i.e., the
aluminum ring plus an effective Josephson junction). This nonlinearity is
essential for a breakdown of the superposition principle for
electromagnetic fields in a vacuum, which permits the detection of the vector
potential arising from the ferromagnetic core of the configuration shown in
Figure 1. Note that we have replaced the red cylindrical shell of Figure 1
by a red ring in the form of a thin, flat circular gasket which has a gap in
it, as shown in Figure 2(b), so as to be able to incorporate the
Meissner-shielded ferromagnetic core into its center during the assembly
process.

One justification\ for using this SQUID-like model \cite{Bloch}\cite%
{Zimmerman}\cite{Jackel}\ goes as follows: The ring gasket of aluminum in
Figure 2(a) must have a large enough gap cut into it in order to enable the
ferromagnetic torus to slip by the gap and enter into the center of the SC
ring, in order to realize the configuration of Figure 1. But in order to
form a complete SC circuit, the gap in the aluminum gasket will be filled in
by pressing onto the top of the gap an overlapping rectangular
aluminum plate (shown in pink in Figure 2(b)), which covers the gap. Thus a
topological \textquotedblleft link\textquotedblright\ can be formed between
a \textquotedblleft ring\textquotedblright\ and a \textquotedblleft
torus\textquotedblright \cite{ChiaoMayArXiv}, with the same topology as that in Figure 1.

Two small screws, which are drilled into the two overlap areas between the
rectangular plate and the gapped gasket, will be used to press tightly the
flat, rectangular aluminum plate onto the gapped region of the flat aluminum
gasket, and thus to complete a SC circuit. However, since the aluminum
material of the ring and of the plate will naturally form an oxide (Al$_{2}$O%
$_{3}$) layer on its surfaces, two oxide-layer Josephson junctions will
naturally be formed at the two overlap areas between the plate and the ring,
after the rectangular plate has been pressed tightly onto the gapped ring by
means of these two small screws. The resulting two in-series, oxide-barrier
Josephson junctions of Figure 2(b) can be thought of as being a
single, effective Josephson junction interrupting the ring. This effective
Josephson junction will obey the following sinusoidal relationship \cite{Bloch}:%
\begin{equation}
I=I_{{\rm J}}\sin \phi   \label{Josephson sinusoidal relationship}
\end{equation}%
where $I$ is the supercurrent flowing through the junction, $I_{{\rm J}}$
is the effective Josephson critical current, and $\phi $ is the phase
difference across the junction, which, for configuration shown in Figure
2(a), is given by the Aharonov-Bohm phase%
\begin{equation}
\phi =\frac{q}{\hbar }\oint {\bf A}\cdot d{\bf l}=2\pi \frac{\Phi }{\Phi _{0}}\propto
\Phi 
\end{equation}%
where $\Phi $ is the total flux enclosed by the ring and%
\begin{equation}
\Phi _{0}=\frac{h}{2e}  \label{flux quantum}
\end{equation}%
is the quantum of flux. Thus the phase shift $\phi $ is directly
proportional to the \emph{total} flux $\Phi $ enclosed by the ring. Note
that this implies that there exists a \emph{nonlinear} relationship exists
between the supercurrent $I$ and the total flux $\Phi $, on account of the
sinusoidal nature of (\ref{Josephson sinusoidal relationship}).

Hysteresis in the ring can arise from the self-inductance of the ring as
follows: The total flux enclosed by the ring will be sum of two terms%
\begin{equation}
\Phi =\Phi _{H}+\Phi _{I}
\end{equation}%
where $\Phi _{H}$ is the flux imposed onto the ring from a pair of Helmholtz
coils, i.e., by an externally applied $H$ field, and $\Phi _{I}$ is the flux
generated by the self-inductance of the ring due to the supercurrent $I$.
For simplicity, we shall first consider the case where the width of the ring
is less than the effective penetration depth, and where we are imposing an
external field close to the effective critical field $H_{{\rm c}}$, so that the
exterior, imposed flux can freely flow in or out from the interior of the
ring near \textquotedblleft criticality\textquotedblright .

Now the flux generated by the Josephson supercurrent $I$ via the inductance $%
L$ of the ring obeys the relationship%
\begin{equation}
\Phi _{I}=LI
\end{equation}%
where $L$ is the self-inductance of the ring. Putting this together with (%
\ref{Josephson sinusoidal relationship}), one obtains the transcendental
equation%
\begin{equation}
I=I_{{\rm J}}\sin \left\{ \frac{2\pi }{\Phi _{0}}\left( \Phi _{H}+LI\right)
\right\}  \label{transcendental equation}
\end{equation}%
One can express this equation as two equations in two unknowns, $\Phi $ and $%
I$, as follows:%
\begin{eqnarray}
I &=&I_{{\rm J}}\sin \left( \frac{2\pi }{\Phi _{0}}\Phi \right)
\label{I vs. Phi} \\
\Phi &=&\Phi _{H}+LI
\end{eqnarray}%
One can rewrite the second equation as the following linear equation for $I$
in terms of $\Phi $:%
\begin{equation}
I=\frac{1}{L}\left( \Phi -\Phi _{H}\right)
\end{equation}%
which has the form of a straight-line relationship%
\begin{equation}
I=a\Phi +b  \label{straight line relationship}
\end{equation}%
where%
\begin{equation}
a=\frac{1}{L}
\end{equation}
and
\begin{equation}
b=-\frac{\Phi _{H}}{L}
\end{equation}%
are the slope and the intercept of the straight-line relationship (\ref%
{straight line relationship}), respectively.

A graphical method of solution to the transcendental equation expressed as
the two equations (\ref{I vs. Phi}) and (\ref{straight line relationship})
is illustrated in Figure 3.

\begin{figure}
\includegraphics[width=4.75in]{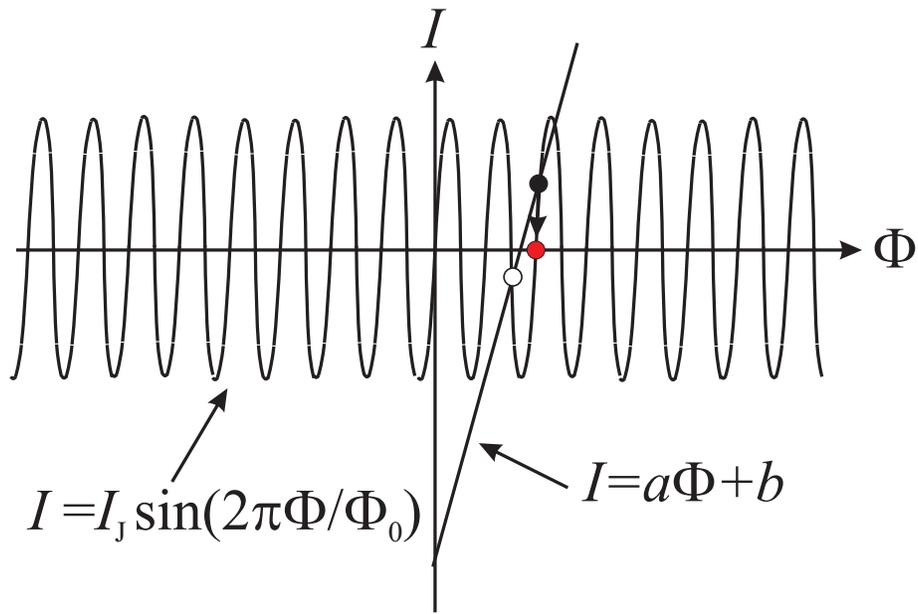}
\caption{Graphical method (not to scale) for the solution of the transcendental
equation (\protect\ref{transcendental equation}), written as the two
equations (\protect\ref{1st graphical equation}) and (\protect\ref{2nd
graphical equation}). The intercept at the black dot represents a stable
solution, and that at the white dot an unstable solution. The red dot
represents the nearest possible metastable flux-trapping state (i.e., a
persistent-current state), which possesses an integer multiple of flux
quanta $\Phi _{0}$ trapped within the ring, i.e., a nonzero remnant
magnetization when $H\rightarrow 0$. Since $\Phi _{0}$ is a very small
quantity, the sinusoidal oscillations of $I$ are very rapid, and the
flux-trapping states are highly metastable.}
\end{figure}

The intercepts\ (the black and the white dots) between the two graphs%
\begin{eqnarray}
I &=&I_{{\rm J}}\sin \left( \frac{2\pi }{\Phi _{0}}\Phi \right) {\rm  and}
\label{1st graphical equation} \\
I &=&a\Phi +b  \label{2nd graphical equation}
\end{eqnarray}%
represent the graphical solutions to the transcendental equation. The white
dot presents an unstable solution (one with an \textquotedblleft
anti-Lenz\textquotedblright\ Faraday-like law). The black dot solution will
eventually settle down to the nearest possible metastable, persistent
current solution, which is represented by the red-dot solution, after the
applied magnetic field is reduced to zero, and a quantized number of flux
lines corresponding to the red-dot solution, has been trapped inside the
ring. This happens when one reduces the external field to values well below
the effective critical field $H_{{\rm c}}$, where the effective penetration depth
becomes much smaller than the width of the red ring. As one further reduces
the applied field until it reaches zero, i.e., as $H\rightarrow 0$, the
remnant field as measured by the Hall probe in Figure 1 will approximately be the trapped
flux corresponding to the red-dot solution, divided by the area of the ring.
(However, the above analysis is for a one-dimensional ring, and needs to be
generalized to a three-dimensional ring in order to confirm the above
conclusions. See Appendix A.)

\begin{figure}
\includegraphics[width=4.75in]{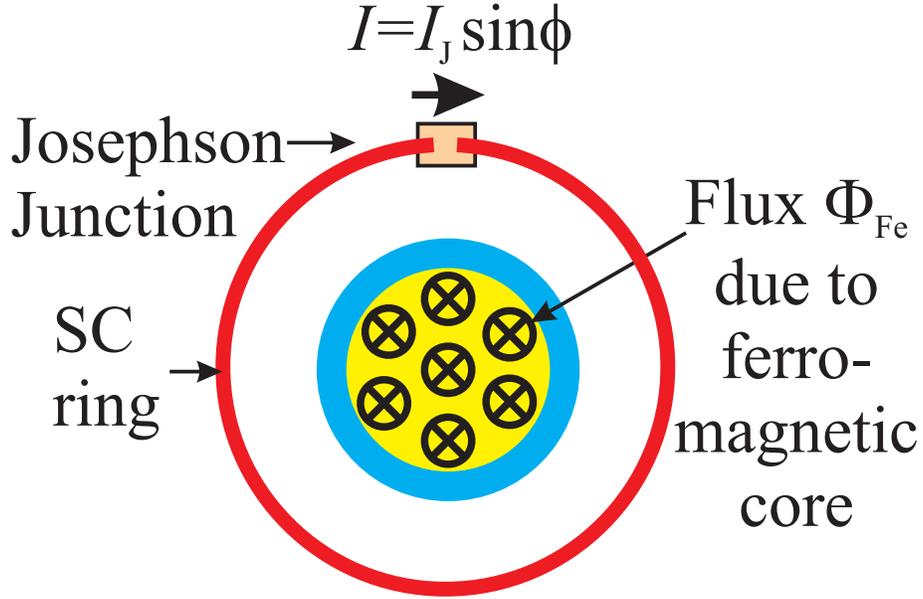}
\caption{SQUID-like model of Figure 2(a)
after the insertion of a ferromagnetic core (yellow) overcoated by a thick
SC coating of lead (blue) into its center (not to scale). The flux from the
ferromagnetic core contributes to the flux seen by the Josephson junction
(pink), even though only the vector potential, but not the ferromagnetic
field from the core, reaches the SC aluminum ring (red).}
\end{figure}

The way to extend the above ideas to the configuration in Figure 1 where
there is a ferromagnetic core thickly overcoated by a strong SC such as lead
or lead-tin solder, is obvious from an inspection of Figure 4. The above
analysis is changed by adding the flux $\Phi _{{\rm Fe}}$ arising from the
ferromagnetic core to the total flux seen by the Josephson junction, i.e.,%
\begin{equation}
\Phi =\Phi _{H}+\Phi _{I}+\Phi _{{\rm Fe}}
\end{equation}%
This has the effect of adding a DC bias to the flux in the graphical
solution of Figure 3 by an amount $\Phi _{{\rm Fe}}$, thus either
increasing\ or decreasing, depending on the sign of $\Phi _{{\rm Fe}}$, the
remnant magnetic field as measured by the Hall probe in Figure 1. This analysis
also extends to the case of a ferromagnetic torus thickly coated with a
strong SC (e.g., lead)\ linked with a ring composed of a weak SC (e.g.,
aluminum or tin), as pictured in the first Figure of \cite{ChiaoMayArXiv}%
.

\section{Appendix A: Case of a wide aluminum ring.}

An external magnetic field $H$ imposed upon the \emph{wide} aluminum ring in
Figure 2 will lead to an inner current $I_{{\rm inner}}$\ flowing along the
inner perimeter of the ring within a penetration depth, and an outer current 
$I_{{\rm outer}}$ flowing along the outer perimeter of the ring within a
penetration depth. These two currents will join together at the effective
Josephson junction to satisfy the Josephson sinusoidal relationship%
\begin{equation}
I=I_{{\rm inner}}+I_{{\rm outer}}=I_{{\rm J}}\sin \phi 
\label{Sine function of total current}
\end{equation}%
The metastable solutions corresponding to flux quantization in integer
multiples of the flux quantum, such as the case of the red-dot solution in
Figure 3, corresponds to solutions with the discrete Aharonov-Bohm phase
values%
\begin{equation}
\phi _{n}=2\pi n,{\rm  }(n=0,\pm 1,\pm 2,...)
\end{equation}%
i.e., the zero-crossings of the sine function (\ref{Sine function of total
current}), or to the quantized flux values%
\begin{equation}
\Phi _{n}=n\frac{h}{2e},{\rm  }(n=0,\pm 1,\pm 2,...)
\end{equation}%
The free energy of the system will have \emph{extremely deep, local minima}
at these quantized values of the flux, which correspond to the \emph{%
extremely metastable} solutions for the wide ring at the effective critical
values of the applied magnetic field (i.e., at \textquotedblleft
criticality\textquotedblright ) where%
\begin{equation}
I_{n}=\left. I_{{\rm inner}}\right\vert _{n}+\left. I_{{\rm outer}%
}\right\vert _{n}=I_{{\rm J}}\sin \phi _{n}=0
\end{equation}%
In other words, at \textquotedblleft criticality\textquotedblright , the
inner and outer currents will satisfy the relationship%
\begin{equation}
\left. I_{{\rm inner}}\right\vert _{n}=-\left. I_{{\rm outer}}\right\vert
_{n}
\end{equation}%
However, as the externally applied magnetic field $H$ is subsequently reduced to
zero, the inner current must remain constant, since the interior flux must
remain quantized, whilst the outer current will be reduced to zero along
with the vanishing exterior $H$ field, i.e.,%
\begin{equation}
\left. I_{{\rm inner}}\right\vert _{n}\rightarrow \frac{\Phi _{n}}{L}{\rm }
\end{equation}
as $H\rightarrow 0$, but
\begin{equation}
\left. I_{{\rm outer}}\right\vert
_{n}\rightarrow 0
\end{equation}%
as $H\rightarrow 0$. Thus there will result a \textquotedblleft remnant\textquotedblright\ field
contained within the inner perimeter of the ring, which will approach the
value%
\begin{equation}
B_{{\rm remnant}}\rightarrow \frac{\Phi _{n}}{\mathcal{A}}
\end{equation}%
as $H\rightarrow 0$,
where $\mathcal{A}$ is the inner area of ring (excluding the cross-sectional area of the
torus), which can be measured by the Hall probe (orange) in Figure 1. This
leads to the possibility of an observation of the \textquotedblleft
asymmetric hysteresis\textquotedblright\ phenomenon in the linked torus-ring
system, as described in \cite{ChiaoMayArXiv}. 

In summary, the flux measured in the Josephson effect in a SQUID-like configuration with a ferromagnetic core is the 
\emph{total} flux enclosed by this kind of SQUID, including the flux from the ferromagnetic core, as
illustrated in Figure 4. It is this that leads to the macroscopic,
Aharonov-Bohm-like effect described here.


\bigskip 

\begin{thebibliography}{9}
\bibitem{ChiaoMayArXiv} R.Y. Chiao, \textquotedblleft Hysteretic method for
measuring the flux trapped within the core of a superconducting lead-coated
ferromagnetic torus by a linked superconducting tin ring, in a novel
Aharonov-Bohm-like effect based on the Feynman path-integral
principle\textquotedblright , arXiv:1205.6029.

\bibitem{Tonomura} A.
Tonomura, N. Osakabe,T. Matsuda, T. Kawasaki, J. Endo, S. Yano, and H.
Yamada, ``Evidence for Aharonov-Bohm effect with magnetic
field completely shielded from electron wave," Phys. Rev.
Lett. {\bf 56}, 792 (1986); A. Tonomura, ``New results on
the Aharonov-Bohm effect with electron interferometry,"
Physica B {\bf 151}, 206 (1988).

\bibitem{Bloch} Another justification for the Josephson sinusoidal
relationship (\ref{Josephson sinusoidal relationship}) for the thin red SC
rings in Figures 1 and 2, is a more general one based solely on the gauge
symmetry considerations given by F. Bloch, which starts from the fact that
the free energy $F$ of the thin ring (i.e., thin compared to the penetration
depth), like that of any homogeneous many-electron system with an
off-diagonal long-range order, must be a periodic function of the flux
enclosed by the ring with a period of quantum of flux $\Phi _{0}\ $given by (%
\ref{flux quantum}), i.e.,%
\begin{equation}
F\left( \Phi +n\Phi _{0}\right) =F\left( \Phi \right), {\rm  }(n=0,\pm
1,\pm 2,...)
\end{equation}%
This follows from the principle of local gauge invariance and the
Aharonov-Bohm effect. Due to the inversion symmetry of space, the free energy%
\begin{equation}
F\left( -\Phi \right) =+F\left( +\Phi \right) 
\end{equation}%
must be an even function of the flux $\Phi $. Since the free energy $F$ is
related to the power delivered to the ring by%
\begin{equation}
\frac{dF}{dt}=IV=I\left( -\frac{d\Phi }{dt}\right) 
\end{equation}%
the supercurrent $I$ induced in the ring is related to the free energy by 
\begin{equation}
I=-\frac{dF}{d\Phi }
\end{equation}%
Since $F$ is a \emph{periodic, even} function of the flux $\Phi $, the
supercurrent $I$ must a \emph{periodic}, \emph{odd} function of the flux $%
\Phi $, i.e.,%
\begin{eqnarray}
I\left( \Phi +n\Phi _{0}\right)  &=&I\left( \Phi \right),  {\rm  }(n=0,\pm
1,\pm 2,...)  \label{I is periodic} \\
I\left( -\Phi \right)  &=&-I\left( +\Phi \right)   \label{I is odd}
\end{eqnarray}%
A Fourier series expansion of any periodic function, such as (\ref{I is
periodic}), with an odd parity symmetry, such as (\ref{I is odd}), must have
as its leading term (i.e., as the fundamental harmonic term)%
\begin{equation}
I=I_{0}\sin \left( 2\pi \frac{\Phi }{\Phi _{0}}\right) 
\label{Bloch's sinusoidal relationship}
\end{equation}%
with some unknown Fourier expansion coefficient $I_{0}$, which must be
determined by experiment. Bloch's sinusoidal relationship (\ref{Bloch's
sinusoidal relationship}) is another form of the Josephson relationship (\ref%
{Josephson sinusoidal relationship}). See F. Bloch, \textquotedblleft
Off-diagonal long-range order and persistent currents in a hollow
cylinder\textquotedblright , Phys. Rev. {\bf 137}, A787 (1965);
\textquotedblleft Simple interpretation of the Josephson
effect\textquotedblright , Phys. Rev. Lett. {\bf 21}, 1241 (1968);
\textquotedblleft Josephson effect in a superconducting
ring\textquotedblright , Phys. Rev. {\bf B2}, 109 (1970).

\bibitem{Zimmerman} A.H. Silver and J.E. Zimmerman, \textquotedblleft
Quantum states and transitions in weakly connected superconducting
rings\textquotedblright , Phys. Rev. {\bf 157}, 317 (1967).

\bibitem{Jackel} J.D. Jackel, R.A. Buhrman, and W.W. Webb, \textquotedblleft
Direct measurement of current-phase relations in superconducting weak
links\textquotedblright , Phys. Rev. {\bf B10}, 2782 (1974).
\end{thebibliography}
\end{document}